\documentclass{aa}  

\usepackage{graphicx}
\usepackage{xcolor}
\usepackage{braket}
\usepackage{bm}
\usepackage{txfonts}
\usepackage{hyperref}
\begin{document}

   \title{Automated calibration of simulated galaxy catalogues for cosmological analyses}

   \author{I. Tutusaus\inst{1,2,3}\fnmsep\thanks{isaac.tutusaus@irap.omp.eu}
   \and
   P. Fosalba\inst{2,3}
   \and
   L. Blot\inst{4,5}
   \and
   P. Tallada-Crespí\inst{6,7}
   \and
   J. Carretero\inst{6,7}
   \and
   F. J. Castander\inst{2,3}
   \and
   E. J. Gonzalez\inst{8,7}
   \and
   A. Alarcon\inst{2}
}

   \institute{Institut de Recherche en Astrophysique et Plan\'etologie (IRAP), Universit\'e de Toulouse,
CNRS, UPS, CNES, 14 Av. Edouard Belin, 31400 Toulouse, France
         \and
             Institute of Space Sciences (ICE, CSIC), Campus UAB, Carrer de Can Magrans, s/n, 08193 Barcelona, Spain
        \and
            Institut d'Estudis Espacials de Catalunya (IEEC), Edifici RDIT, Campus UPC, 08860 Castelldefels (Barcelona), Spain
        \and
            Center for Data-Driven Discovery, Kavli IPMU (WPI), UTIAS, The University of Tokyo, Kashiwa, Chiba 277-8583, Japan
        \and
            Laboratoire d’étude de l’Univers et des phénomènes eXtrêmes, Observatoire de Paris, Université PSL, Sorbonne Université, CNRS, 92190 Meudon, France
        \and
            Centro de Investigaciones Energ\'eticas, Medioambientales y Tecnol\'ogicas (CIEMAT), Avenida Complutense 40, 28040 Madrid, Spain
        \and
            Port d'Informaci\'{o} Cient\'{i}fica, Campus UAB, C. Albareda s/n, 08193 Bellaterra (Barcelona), Spain
        \and
            Institut de F\'{i}sica d'Altes Energies (IFAE), The Barcelona Institute of Science and Technology, Campus UAB, 08193 Bellaterra (Barcelona), Spain
             }

   \date{Received XXX; accepted YYY}

 
  \abstract
{Simulated galaxy catalogues have become an essential tool for preparing and exploiting observations from galaxy surveys. They constitute a key ingredient in modelling the systematic uncertainties present in the analysis. However, in order to reach the large volume and high precision required for galaxy surveys, we generally populate dark matter haloes with galaxies following certain theoretical recipes. Such recipes contain free parameters that are calibrated comparing the simulations against observations, but the creation of galaxy mocks is a stochastic process with a large number of free parameters to calibrate. We present a new pipeline, based on the differential evolution algorithm, that can calibrate galaxy mocks in a fully automated way for realistic scenarios with a large parameter space. We apply the pipeline to galaxy mocks built on a combination of halo occupation distribution and sub-halo abundance matching techniques. We show that our pipeline can properly calibrate the galaxy mocks against observations for both $\Lambda$CDM and modified gravity halo catalogues. This type of calibration pipeline provides a new tool for automating the calibration of future massive galaxy mocks.}


   \maketitle
%

\section{Introduction}
Galaxy surveys constitute one of the most powerful ways to constrain the plethora of available cosmological models and try to better understand the accelerated expanding Universe in which we live. This claim is supported by recent results from galaxy survey observations such as the Dark Energy Survey\,\citep{2022PhRvD.105b3520A}, the Kilo-Degree Survey\,\citep{2021A&A...646A.140H}, or the Dark Energy Spectroscopic Instrument\,\citep{2024arXiv240403002D}, which provide exquisite constraints on the main parameters of the $\Lambda$CDM concordance model. However, to fully exploit and interpret the observations provided by galaxy surveys, precise cosmological simulations are required. Simulations are not only considered to design galaxy surveys, but they are also used to better understand selection effects, calibrate systematic uncertainties, or inform on the scales that can be trusted with our modelling. Given the key role of simulations in the final analysis, significant efforts have been made in different collaborations to build this kind of simulations\,\citep[see e.g.,][]{DeRose2021,Smith2020,Alam2021,Castander2024}. 

Galaxy surveys can probe very large volumes of the Universe down to very small scales. This implies that the required simulations should cover huge volumes with very high resolution, making them computationally expensive. The standard approach consists in building very large N-body dark matter simulations, identify the dark matter haloes through some halo finding algorithm, and then populate these haloes with galaxies following different theoretical recipes. The connection between dark matter haloes and galaxies is usually done through hydro-dynamical simulations\,\citep[see e.g., the review in][]{Springel2010}, semi-analytic models\,\citep{Benson2012}, and extensions of the halo model, like the halo occupation distribution (HOD) or the sub-halo abundance matching (SHAM) technique\,\citep[for a review, see e.g.,][]{2018ARA&A..56..435W}. 

All the different recipes to implement the connection between dark matter haloes and galaxies have free nuisance parameters that allow us to change the properties of the galaxy mock catalogue being generated. In order to have useful mocks for cosmological analysis, we need to ensure that the values of these parameters are set in such a way that the catalogue properties match the observations. The selection of the free parameters might seem an easy task when considering only a few galaxy properties (and thus a few parameters). However, current galaxy mocks should accurately reproduce the most important global properties of different galaxy populations (like the luminosity function or the colour distributions) and also how their clustering behaves as a function of these quantities over a wide dynamical range (e.g., from linear to highly non-linear scales). This kind of simulations leads us to a scenario in which several tens of parameters need to be determined.

In this article we aim at providing an automated pipeline to determine the values of the free parameters connecting galaxies and dark matter haloes. Our goal is to be able to calibrate the galaxy mocks against real observations in a fully automated way, for realistic cases (more than 20 dimensions in the calibration). To do this, we consider the halo catalogues provided in \citet{Arnold2019}, we populate the haloes with galaxies following the combination of HOD and SHAM techniques presented in \citet{Carretero2015}, and we incorporate a calibration pipeline that stochastically minimizes the discrepancies between the galaxy mocks and observational data. The article is organised in the following way: we describe the recipes followed to connect dark matter haloes and galaxies in Sect.\,\ref{sec:mocks}. We present the calibration pipeline of this work in Sect.\,\ref{sec:calibration}, we briefly discuss the production of the simulations in Sect.\,\ref{sec:production}, and we show the main results in Sect.\,\ref{sec:results} before concluding in Sect.\,\ref{sec:conclusions}.

\section{Galaxy mock catalogues}\label{sec:mocks}
In this section we provide a brief description of how the galaxy mocks are generated starting from the halo catalogues. We follow closely the approach described in \citet{Carretero2015} that was first used to populate the Grand Challenge run of the MICE simulations\,\citep{Fosalba2013a,Crocce2013,Fosalba2013b} and later adapted for the {\it Euclid} Flagship galaxy mock\,\citep{Castander2024}. Therefore, we refer the reader to \citet{Carretero2015} for more details on the specific procedure. 

\subsection{Modelling the halo -- galaxy connection}
To properly reproduce the observations, a model connecting haloes and galaxies is required. In this work we use an HOD model combined with the abundance matching (AM) technique to assign galaxies to the different haloes.

We start with a simple HOD form in which each halo may contain a central galaxy (depending on whether the mass is above a certain threshold) and the average number of satellite galaxies is given by a power law of the halo mass. Our HOD is parametrised with three free parameters: $M_{\rm min}$, the minimum halo mass needed to host a central galaxy; $M_1$, the halo mass for which a halo contains, on average, one satellite galaxy; and $\alpha$, the slope of the satellite mean occupation. We can therefore write the mean number of central galaxies as
\begin{equation}
    \braket{N_{\rm cen}}=1\,,\text{ if}\,M_{\rm h}\geq M_{\rm min}\,,
\end{equation}
where $M_{\rm h}$ is the mass of the halo. We assign the number of satellite galaxies using a Poisson distribution with mean
\begin{equation}\label{eq:nsat}
    \braket{N_{\rm sat}}=\left(\frac{M_{\rm h}}{M_1}\right)^{\alpha}\,,\text{ if}\,M_{\rm h}\geq M_{\rm min}\,.
\end{equation}

However, this HOD form is too simple to reproduce observations for different luminosity threshold samples with a single set of parameters. Therefore, we further model the dependence of $M_1$ on $M_{\rm min}$ and the mass of the halo following \citet{Carretero2015} as
\begin{align}
    M_1=M_{\rm min}\times \{&\left[(a_1-a_2)\,\text{tanh}(s_1(b_1-\,\text{log}(M_{\rm h})))\right.\nonumber\\
    &+(a_3-a_2)\,\text{tanh}(s_2(\text{log}(M_{\rm h})-b_2))\nonumber\\
    &+(a_1+a_3)]/2\}\,,
\end{align}
where $a_i,b_j,s_k$ are free parameters for $i=1,2,3$ and $j,k=1,2$. The $a_i$ values provide the correction to $M_{\rm min}$ for different values of $M_{\rm h}$, while $b_j$ give the values of $M_{\rm h}$ at which the transition between different amplitudes should occur, and $s_k$ denote the slope of such transitions.

Given the HOD model and the halo mass function of the halo catalogue, d$n/$d$M_{\rm h}$, we can compute the cumulative mean number density of galaxies inhabiting haloes splitting them between centrals and satellites
\begin{equation}
    n_{\rm gal}(> M_{\rm min})=n_{\rm cen}(>M_{\rm min})+n_{\rm sat}(>M_{\rm min})\,,
\end{equation}
with
\begin{equation}\label{eq6}
    n_{\rm cen}(>M_{\rm min})=\int_{M_{\rm min}}^{\infty}N_{\rm cen}(M_{\rm h})\frac{\text{d}n}{\text{d}M_{\rm h}}\,\text{d}M_{\rm h}\,,
\end{equation}
and
\begin{equation}
    n_{\rm sat}(>M_{\rm min})=\int_{M_{\rm min}}^{\infty}N_{\rm sat}(M_{\rm h})\frac{\text{d}n}{\text{d}M_{\rm h}}\,\text{d}M_{\rm h}\,.
\end{equation}

On the other hand, using the best fit of the modified Schechter function derived from observations\,\citep{Blanton2003b,Blanton2005b}, we compute the mean number density of galaxies with luminosities brighter than a certain threshold
\begin{equation}
    n_{\rm gal}(>L_{\rm r})=\int_{L_{\rm r}}^{\infty}\frac{\text{d}n}{\text{d}L_{\rm r}}\,\text{d}L_{\rm r}\,,
\end{equation}
where $L_{\rm r}$ stands for a luminosity threshold in the r band.

Using the SHAM technique, we derive a relation between the mass of the halo, $M_{\rm h}$, and the luminosity of the galaxy, $L_{\rm gal}$, equalling both cumulative functions
\begin{equation}
    n_{\rm gal}(>M_{\rm min})=n_{\rm gal}(>L_{\rm r})\,.
\end{equation}

We assume that central galaxies follow this $M_{\rm h}$ -- $L_{\rm gal}$ relation and assign their luminosity depending on the mass of the haloes they populate. We then assign luminosities to satellite galaxies drawing them from their cumulative luminosity function and imposing that they cannot be much brighter than the central galaxy of the halo: $L_{\rm sat}\leq 1.05 L_{\rm cen}$. We note that this exact value is somewhat arbitrary, but we have verified that it is enough to allow for a proper calibration of the luminosity function, as it will be discussed below.

Because of the SHAM technique described above, the most luminous galaxies end up in the most massive haloes. It turns out that, following this technique, the most luminous galaxies are more clustered than observations, independently of the HOD used. Therefore, we introduce a scatter in the halo mass -- galaxy luminosity relation. In more detail, we follow \citet{Carretero2015} in computing an unscattered luminosity function, $\Phi (\text{log}\,L)_{\rm unscat}$, that results in a cumulative luminosity function in agreement with observations, $\Phi (\text{log}\,L)_{\rm obs}$, once the scatter is introduced. This is essentially a deconvolution problem given by

\begin{equation}
    \Phi (\text{log}\, L)_{\rm obs} = \int_{-\infty}^{\infty} \Phi (\text{log}\,L')_{\rm unscat}\, G(\text{log}\,L'-\text{log}\,L)\,\text{d}\,\text{log}\,L'\,,
\end{equation}
where $G$ stands for a Gaussian distribution with scatter $\sigma_{\text{log}\,L}$. We consider this scatter as one of our free parameters to calibrate. We also follow \citet{Carretero2015} in applying this scatter only above a certain luminosity threshold $L_{\rm AM}$, which we also consider a free parameter to calibrate.

Finally, when assigning luminosities to satellite galaxies we model the cumulative luminosity function of satellite galaxies within each halo using a modified four-parameter Schechter function\,\citep[see Eq. 13 in][]{Castander2024}. We further parametrise the characteristic luminosity of satellite galaxies, $L_*$, as a function of the luminosity of the central galaxy, $L_{\rm cen}$, as

\begin{equation}\label{eq:lstar}
    \text{log}(L_{*})= \text{log}(L_{\rm cen})-\left\{a_{\rm AM}+b_{\rm AM}\left[\text{log}(L_{\rm cen})-c_{\rm AM}\right]\right\}\,,
\end{equation}
where $a_{\rm AM}, b_{\rm AM}$, and $c_{\rm AM}$ are free parameters to calibrate.

\subsection{Galaxy positions and velocities}
Once the relation between halo mass and galaxy luminosity has been established, we start placing central galaxies at the centre of their host haloes. Satellite galaxies are then distributed according to a spherical Navarro-Frenk-White\,\citep[NFW,][]{NFWref} profile
\begin{equation}
    \frac{\rho(r)}{\bar{\rho}}=\frac{\Delta_{\rm vir}(z)}{3\Omega(z)}\frac{c^3f(c)}{x(1+x)^2}\,,
\end{equation}
where $x\equiv c(M_{\rm h})r/r_{\rm vir}$, $c$ is the concentration parameter, $\bar{\rho}$ is the average density of the background universe, $\Delta_{\rm vir}/\Omega$ is the average density within the virial radius, $f(c)=[\text{log}(1+c)-c/(1+c)]^{-1}$ and $\Delta_{\rm vir}(z)=18\pi^2+82[\Omega(z)-1]-39[\Omega(z)-1]^2$\,. We note that we choose a spherical NFW for simplicity. It could be easily generalised to more realistic triaxial NFW haloes, as it was done in the MICE simulations\,\citep{Carretero2015}.

We consider both a $\Lambda$CDM and an $f(R)$ modified gravity mock to test our calibration pipeline. In both cases, we compute the concentration parameter $c$ using \citep{springel2008}
\begin{align}
\delta_c &= 7.213\, \delta_V = 7.213 \times 2 \left( \frac{v_{\rm max}}{H_0\, r_{\rm max}}\right)^2\,, \nonumber\\ \delta_c &= \frac{200}{3} \frac{c^3}{\log(1+c) - c / (1+c)}\,,\label{concentration}
\end{align}
where $v_{\rm max}$ and $r_{\rm max}$ are the maximum circular velocity and the corresponding radius. We note that in practice the concentration parameters will be different due to the different $v_{\rm max}$ and $r_{\rm max}$ between the two gravity models. We further compute the halo radius as
\begin{equation}
r_{\rm h}=\left(\frac{3\, M_{\rm fof}}{4\pi\, 200\, \rho_{\rm crit}}\right)^{1/3}\,,
\end{equation}
where $M_{\rm fof}$ is the friends-of-friends halo mass, $\rho_{\rm crit}$ is the critical energy density, and we assume that $r_{\rm h}\approx r_{200,\rm crit}$.

Given the number of satellites in each halo provided by the HOD parametrisation, we draw samples of the density distribution to obtain the radial position of the satellite galaxies. We assume that haloes are spherical and draw randomly two position angles that keep the surface density uniform in a sphere.

According to \citet{Carretero2015} and \citet{Watson2012}, satellite galaxies need to be more concentrated than their underlying dark matter NFW density profile. If not, the clustering of galaxies in the 1-halo regime (where satellite galaxies are more relevant) is systematically below the observations. Therefore, we follow \citet{Carretero2015} in applying a factor $\mathcal{C}$ to reduce the distance and place satellite galaxies closer to the centre of the halo 
\begin{equation}
    \mathcal{C}=\left\{
    \begin{array}{ll}
        0.3\,, & \text{ if } M_{\rm r}>-19.25\,, \\
        0.3+\frac{0.65-0.3}{-21+19.25}(M_{\rm r}+19.25)\,, & \text{ if }-21\leq M_{\rm r} \leq -19.25\,, \\
        0.65\,, & \text{ if } M_{\rm r} < -21\,,
    \end{array}
    \right.
\end{equation}
where $M_{\rm r}$ is the absolute magnitude in the r band.

With respect to galaxy velocities and starting with central galaxies, we assume them to be at rest at the centre of the halo with no relative velocity. Therefore, their velocity is given by the velocity of the halo. 
For satellite galaxies, instead, we add a perturbation drawn from a Gaussian distribution with width given by the measured halo velocity dispersion. 

\begin{table}
	\centering
	\caption{Summary of the 23 free parameters of the galaxy mock pipeline that enter into the calibration process. See the text for detailed explanations.}
	\label{tab:table1}
    \resizebox{\columnwidth}{!}{%
	\begin{tabular}{ll}
    \hline
    $\alpha$ & Slope of the satellite mean occupation\\
    $a_1,a_2,a_3$ & Corrections to $M_{\rm min}$ for different $M_{\rm h}$ values\\
    $b_1,b_2$ & Transition values of $M_{\rm min}$\\
    $s_1,s_2$ & Slopes of the $M_{\rm min}$ transitions\\
    $\sigma_{\text{log}\,L}$ & Abundance matching scatter\\
    $L_{\rm AM}$ & Abundance matching luminosity threshold\\
    $a_{\rm AM},b_{\rm AM},c_{\rm AM}$ & Parameters of the characteristic -- central luminosity relation\\
    $f^{\rm red}_{i,\rm sat}$, $i=1,\ldots,5$ & Fraction of red satellites in $i$-th luminosity bin\\
    $f^{\rm green}_{i,\rm sat}$, $i=1,\ldots,5$ & Fraction of green satellites in $i$-th luminosity bin\\
		\hline
	\end{tabular}
    }
\end{table}

\subsection{Galaxy colours}
Galaxy colours are basic properties of galaxies that inform us about the details in the process of galaxy formation and evolution and, as such, allow us to classify them using their behaviour in colour space. Because of this, beyond assigning positions and velocities to galaxies, we also want to reproduce the observed clustering of galaxies as a function of their colour.

In order to assign colours to galaxies we assume they can be classified into three populations: blue cloud, red sequence, and green valley. Given the HOD parametrisation described in the previous sections, we first compute the fraction of central, $f_{\rm cen}$, and satellite galaxies, $f_{\rm sat}$, as a function of luminosity. We then consider SDSS data\,\citep{Blanton2003a} to extract the desired fraction of red, green, and blue galaxies ($f^{\rm red},f^{\rm green},f^{\rm blue}$, respectively) as a function of luminosity. We can finally assign the relative fractions of central galaxies in each colour with
\begin{equation}\label{eq:color1}
    f^{\rm red}_{\rm cen}=\frac{f^{\rm red}-f^{\rm red}_{\rm sat}\times f_{\rm sat}}{f_{\rm cen}}\,,
\end{equation}
\begin{equation}\label{eq:color2}
    f^{\rm green}_{\rm cen}=\frac{f^{\rm green}-f^{\rm green}_{\rm sat}\times f_{\rm sat}}{f_{\rm cen}}\,,
\end{equation}
\begin{equation}\label{eq:color3}
    f^{\rm blue}_{\rm cen}=1-f^{\rm red}_{\rm cen}-f^{\rm green}_{\rm cen}\,,
\end{equation}
where the fractions of satellite galaxies $f^{\rm red}_{\rm sat}$ and $f^{\rm green}_{\rm sat}$ are free parameters and $f^{\rm blue}_{\rm sat}=1-f^{\rm red}_{\rm sat}-f^{\rm green}_{\rm sat}$. In our analysis we bin $f^{\rm red}_{\rm sat}$ and $f^{\rm green}_{\rm sat}$ into 5 luminosity bins and allow each bin to vary freely, leading to a total of 10 additional free parameters.

\section{Calibration}\label{sec:calibration}

The pipeline used in this work to generate galaxy mocks contains several free parameters to model the different quantities relevant for the HOD and SHAM techniques, as well as galaxy colours, as detailed in the previous section. These parameters should be varied simultaneously to minimise the difference between the measurements in real data and the measurements in our mocks. In other words, these parameters need to be calibrated. We summarise the 23 free parameters considered in this work in Table\,\ref{tab:table1}.

For this calibration we consider the projected two-point correlation function as our main observable. Therefore, we aim at minimising the discrepancy between the measured projected correlation function, $w_{\rm p}^{\rm obs}$, and the one obtained from the mocks, $w_{\rm p}^{\rm mock}$. This implies calibrating the 23 free parameters to produce the mock. While in the past this could be achieved by trying a few parameter combinations until the agreement was satisfying, given the size of the parameter space for our model this approach would be impractical.

In this work we present a novel technique where we implement a calibration procedure that enables us to sample the parameter space and provide the values giving the best agreement with the data in a fully automated way. The details of this method are presented in the following.

The first step of the calibration consists in sampling a given Latyn hyper-cube\,\footnote{We note that a Latyn hyper-cube is just the initial choice to start the sampling in the parameter space. A regular grid, for example, would eventually provide the same results, since the sampling would also converge to the minimum in this case.} in the parameter space. 
In each point we generate a galaxy mock, measure $w_{\rm p}^{\rm mock}$, and compute the value of the $\chi^2$ between the measurements from the mock and the real measurements
\begin{equation}\label{eq:chi2}
    \chi^2=\left(w_{\rm p}^{\rm obs}- w_{\rm p}^{\rm mock}\right)^TC^{-1}\left(w_{\rm p}^{\rm obs}- w_{\rm p}^{\rm mock}\right)\,,
\end{equation}
where $C$ stands for the covariance matrix of the data.

We note that the approach described until here is not significantly different than a standard Monte Carlo Markov chain, except for the choice of points in the parameter space. However, it is important to mention that each evaluation in a point of this parameter space is computationally very expensive, since it implies generating a galaxy mock and measuring the vector of projected correlation functions on it. Moreover, we are not interested in the posterior of the calibration parameters, but just on their best fit. This is the main input needed to generate mocks close to the real measurements. A simple way forward could be to use a minimisation algorithm to go directly to the minimum of the $\chi^2$ function, like \texttt{Minuit}\,\citep{minuit_ref}. But the production of galaxy mocks contains an intrinsic random component when assigning the position, velocities, and colours of galaxies. This introduces an important stochastic behavior in our problem and makes the standard minimisation algorithms unusable.

To overcome this limitation, we consider the differential evolution stochastic minimisation algorithm first proposed by \citet{Storn1997}. The essential idea of the algorithm is to consider a set (population) of points in the parameter space (called candidate solutions) as the possible minimum. In an iterative way, the candidate solutions are combined to generate a new population and the $\chi^2$ is evaluated for each one of the new candidates. If these are better than the previous candidates (smaller $\chi^2$), they are accepted and belong to the population. Otherwise they are discarded. In the following we provide a more formal description of the methodology.

Let $\bm{x} \in \mathbb{R}^n$ be a candidate solution of the population $\bm{X}$, where $n = 23$ in our case. We first initialize all members of the population using a Latyn hyper-cube sampling and evaluate the $\chi^2$ at each one of them. After this, two members of the population are randomly chosen and their difference is used to mutate the best candidate solution so far
\begin{equation}
    \bm{x}'=\bm{x}_0+M\cdot (\bm{x}_i-\bm{x}_j)\,,
\end{equation}
where $\bm{x}_0$ is the best candidate solution (minimum $\chi^2$ of the population), $M$ is a mutation constant, and $\bm{x}_i$, $\bm{x}_j$ denote two randomly chosen members of the population.

We then build a trial population $\bm{Y}$ using candidates $\bm{x}$ from the original population and mutated candidates $\bm{x}'$. In order to decide whether $\bm{x}$ or $\bm{x}'$ will become a member of the trial population we draw a random number between 0 and 1 using a binomial distribution. If this random number is smaller than a recombination constant $R$ then the new member of $\bm{Y}$ is $\bm{x}'$, otherwise it is given by $\bm{x}$. Once the trial population has been selected we evaluate the $\chi^2$ for each one of the members. If their $\chi^2$ values are smaller than the ones from the initial population $\bm{X}$ we update the members. If the best candidate of the trial population has a smaller $\chi^2$ than the best candidate of $\bm{X}$ we also replace it. Once $\bm{X}$ has been updated with the corresponding members of $\bm{Y}$ we repeat the process. The iterative procedure will end when either the maximum number of evaluations or the requested tolerance is achieved. The latter will occur when the standard deviation of the $\chi^2$ values of the population is smaller than an absolute tolerance $A$ plus a tolerance $T$ times the absolute value of the mean, $\mu$, of the $\chi^2$ values of
the population
\begin{equation}
    \sigma\left(\left\{\chi^2(\bm{x})\right\}\right) \leq A+T\cdot \left|\mu\left(\left\{\chi^2(\bm{x})\right\}\right)\right|\,.
\end{equation}

Using a larger set of members for the population together with higher mutation and lower recombination helps in finding the global minimum, since it widens the search radius when sampling the parameter space. However, it converges slower. In order to increase the
convergence rate we update the new population after evaluating the $\chi^2$ for a given member of the trial population, instead of waiting until the end of the iteration, as it was done in \citet{Wormington1999}. This implies that subsequent new candidates will immediately benefit from the new additions to the population. In practice, we consider the implementation of the differential evolution stochastic minimisation available in \texttt{SciPy}\,\citep{scipy_ref}.

The automated calibration pipeline presented in this work has also been used to calibrate (against observations) the galaxy mocks used in the baryon acoustic oscillations analysis of the Dark Energy Survey Year 3 analysis\,\citep{Ferrero2021}.

\begin{figure}
\begin{center}
    \includegraphics[scale=0.6]{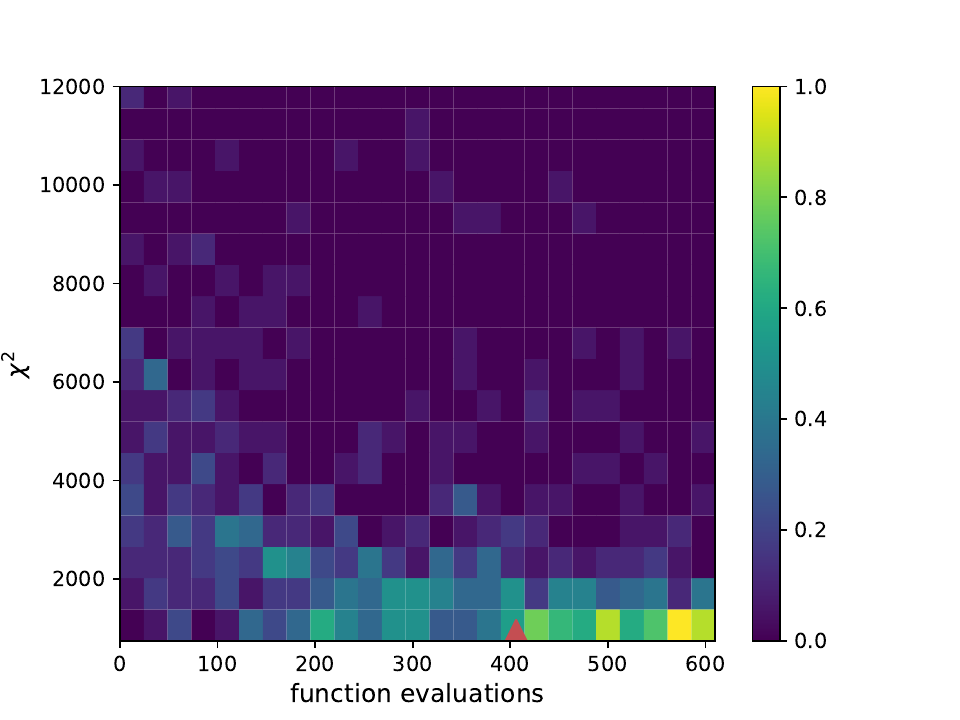}\\
    \includegraphics[scale=0.6]{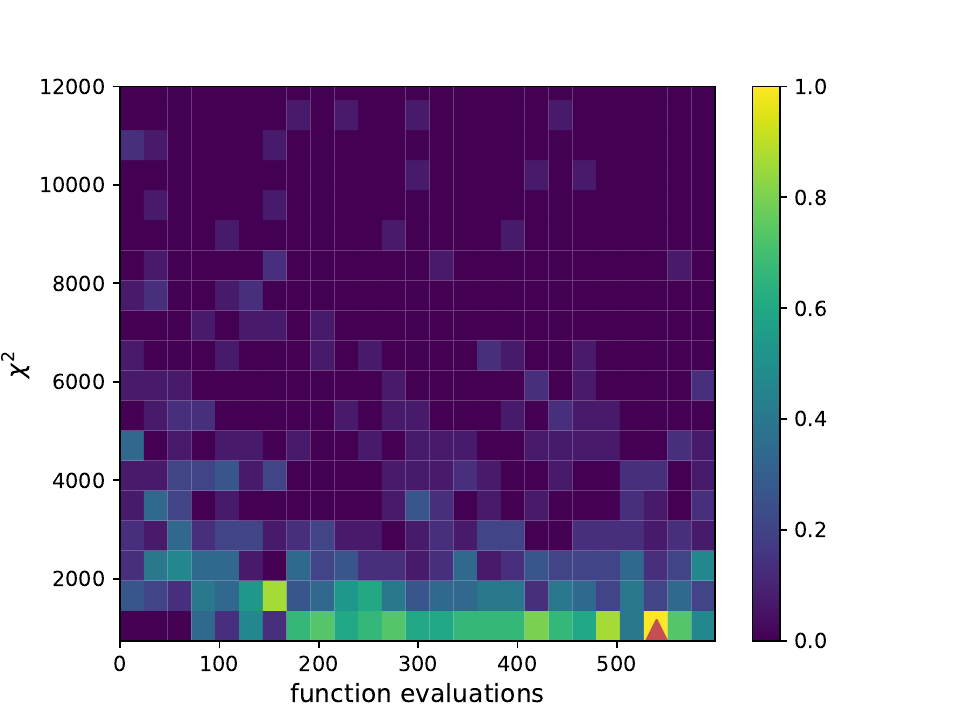}
    \caption{Histograms of the $\chi^2$ value as a function of the number of evaluations. The absolute minimum of the $\chi^2$ is represented by the red triangle. The stochastic behaviour of the problem can be observed, as well as the convergence of $\chi^2$ values as a function of the number of evaluations. {\it Top panel:} $\Lambda$CDM catalogue. {\it Bottom panel:} Modified gravity $f(R)$ catalogue. The colour map corresponds to a linear mapping between 0 and 1, with 0 being the lowest value of the two-dimensional histogram and 1 being its highest value. See the text for more details.}\label{fig:chi2}
\end{center}
\end{figure}

\section{Production}\label{sec:production}
As it has been mentioned in Sect.\,\ref{sec:calibration}, each evaluation of a point in the parameter space is computationally expensive, as it requires generating a galaxy mock and measuring its corresponding projected correlation function.
The mock generation pipeline is composed of a sequence of algorithms that produce different galaxy properties following the recipes described in Sect.\,\ref{sec:mocks}. In order to have an efficient mock production process, the pipeline has been implemented using \texttt{Apache Spark}\,\footnote{\url{https://spark.apache.org}}. This enables the automatic splitting, distribution, and orchestration of all the tasks involved on different nodes.

The algorithm for computing the projected correlation function was also designed ad-hoc to better exploit the locality and parallelism opportunities of the \texttt{Apache Spark} platform. This custom implementation also integrates the calculation of the sample variance using the jackknife resampling technique and takes advantage of a pixelisation algorithm (\texttt{HEALPix}\,\footnote{\url{https://healpix.jpl.nasa.gov/}}) to reduce the number of pairs that have to be processed.

For this work, the mock generation pipeline and the correlation function algorithm have been compacted into a single function. This function takes as arguments all the free parameters that define the models behind each property and returns the final $\chi^2$. This new function has then been minimised with the differential evolution stochastic minimisation described in Sect.\,\ref{sec:calibration}. Once the free parameters have been calibrated, the pipeline has been run again with the corresponding values to generate the final, calibrated, galaxy mocks. All the tasks required for this work have run using the Port d'Informació Científica (PIC)\,\footnote{\url{https://www.pic.es}} HTCondor cluster. The total amount of computing time (including the sum of wall-time and suspension-time) for the calibration of the simulations described in Sect.\,\ref{sec:results} has been of 4801 hours using 130 jobs in parallel for the general relativity mock and 140 jobs for the modified gravity mock. This corresponds to a wall-clock time of 35.6 hours for two catalogues of around 5 million galaxies each below redshift 0.1, which corresponds to the volume used for the calibration.

\section{Results}\label{sec:results}

In this section we illustrate how the calibration procedure presented in Sect.\,\ref{sec:calibration} operates to calibrate a galaxy mock generated according to the methodology presented in Sect.\,\ref{sec:mocks}. However, this procedure can be easily adapted to any other parametric galaxy assignment scheme. We consider the halo catalogues presented in \citet{Arnold2019} and real observations from SDSS at $z=0.1$ to calibrate the galaxy mocks: the luminosity function\,\citep{Blanton2003b,Blanton2005b}, the clustering of galaxies\,\citep{Zehavi2011}, and the colour-distribution as a function of the absolute magnitude in the $r$ band\,\citep{Blanton2005c}. We make use of the CosmoHub platform to access and manipulate the simulations \citep{cosmohub_ref1,cosmohub_ref2}, together with the SciPIC pipeline\,\citep{cosmohub_ref2} to populate the halo catalogues with galaxies.

The authors in \citet{Arnold2019} presented a set of four high-resolution simulations for both $\Lambda$CDM and the modified gravity $f(R)$ cosmology\,\citep{Buchdahl1970,HuSawicki2007} that were run with the \texttt{MG-Gadget} code\,\citep{mggadget}. All simulations included $2048^3$ particles in boxes of 768\,$h^{-1}$Mpc and 1536\,$h^{-1}$Mpc, which yielded mass resolutions of $M_{\rm part}=4.5\times 10^9 \,h^{-1}M_{\odot}$ and $M_{\rm part}=3.6\times 10^{10}\, h^{-1}M_{\odot}$, respectively. In this work we consider the two simulations corresponding to the higher resolution runs for both gravity theories, $\Lambda$CDM and $f(R)$ with $|\bar{f}_{R,0}|=10^{-5}$, to show that our calibration pipeline does not depend on the underlying gravity model. Given the simulations, the authors in \citet{Arnold2019} used the friends-of-friends halo finder of \texttt{P-GADGET3} \citep[a modification of the \texttt{GADGET-2} presented in][]{Springel2005} to obtain halo catalogues. The minimum number of particles for haloes is 32. These halo catalogues constitute the main input of our calibration pipeline.

Once the halo catalogues are ingested into the calibration pipeline, for each set of calibration parameter values (see Table\,\ref{tab:table1}), we generate the galaxy mock according to the recipe presented in Sect.\,\ref{sec:mocks}. We then compute the $\chi^2$ for three different quantities. The first $\chi^2$ corresponds to the agreement between the SDSS luminosity function at $z=0.1$ in the $r$ band\,\citep{Blanton2003b} and the one measured from the corresponding galaxy mock 
\begin{equation}
    \chi^2_{\rm LF}=\frac{(\text{LF}_{\rm obs}-\text{LF}_{\rm mock})^2}{\sigma_{\rm LF}^2}\,,
\end{equation}
where LF stands for luminosity function and we allow for some uncertainty of the luminosity function at high luminosities setting $\sigma_{\rm LF}=10^{-3}$. This value is enough to ensure that the luminosity functions agree very well in a broad range of luminosities, while still allowing for some uncertainty at very bright magnitudes, where the measurements might be less reliable. Small variations to this exact value should have a negligible impact in the final results. Moreover, in case we had access to the uncertainties on the measurements, the pipeline could easily be modified to account for them in the $\chi^2$ computation. We note that we neglect any correlation in the luminosity function between different luminosities and we implicitly assume a summation on the different luminosity bins.

The second $\chi^2$ corresponds to the comparison between the observed and measured (in the mock) projected two-point correlation function of galaxies as a function of luminosity threshold\,\citep{Zehavi2011}
\begin{equation}
    \chi^2_{\rm GC}=\frac{\left(w_{\rm p}^{\rm obs}-w_{\rm p}^{\rm mock}\right)^2}{\left(\sigma_{w_{\rm p}^{\rm obs}}^2+\sigma_{w_{\rm p}^{\rm mock}}^2\right)}\,,
\end{equation}
where GC stands for galaxy clustering and $\sigma$ denotes the uncertainty of the corresponding correlation function. We note that we neglect the correlation between different bins concerning the projected two-point correlation function, because the full covariance results were not provided in \citet{Zehavi2011}. However, the calibration pipeline remains sufficiently flexible to include the covariance in the $\chi^2$ computation in case it is available. Like in the previous case, we assume an implicit summation on the separation bins and the different luminosity thresholds.

The third and last $\chi^2$ corresponds to the comparison of the projected two-point correlation function as a function of luminosity distinguishing between blue and red galaxies\,\citep{Zehavi2011}
\begin{equation}
    \chi^2_{\rm colour}=\left.\frac{\left(w_{\rm P}^{\rm obs}-w_{\rm P}^{\rm mock}\right)^2}{\left(\sigma_{w_{\rm P}^{\rm obs}}^2+\sigma_{w_{\rm P}^{\rm mock}}^2\right)}\right|_{\rm blue}+\left.\frac{\left(w_{\rm P}^{\rm obs}-w_{\rm P}^{\rm mock}\right)^2}{\left(\sigma_{w_{\rm P}^{\rm obs}}^2+\sigma_{w_{\rm P}^{\rm mock}}^2\right)}\right|_{\rm red}\,.
\end{equation}

Once again, we implicitly assume a summation on the separation bins of the projected correlation function and the different luminosity thresholds.

With these three quantities, we minimise the final $\chi^2$ given by
\begin{equation}
    \chi^2=\chi^2_{\rm LF}+\chi^2_{\rm GC}+\chi^2_{\rm colour}\,,
\end{equation}
with the differential minimisation algorithm described in Sect.\,\ref{sec:calibration}. We note that this expression implicitly assumes no correlation between the total projected correlation function and the projected correlation function splitting in blue and red galaxies. No correlation is either considered between the luminosity function and the projected correlation function. However, the methodology presented in this work can also be directly used in cases where such correlations are included.

\begin{figure}[t]
\begin{center}
    \includegraphics[scale=0.38]{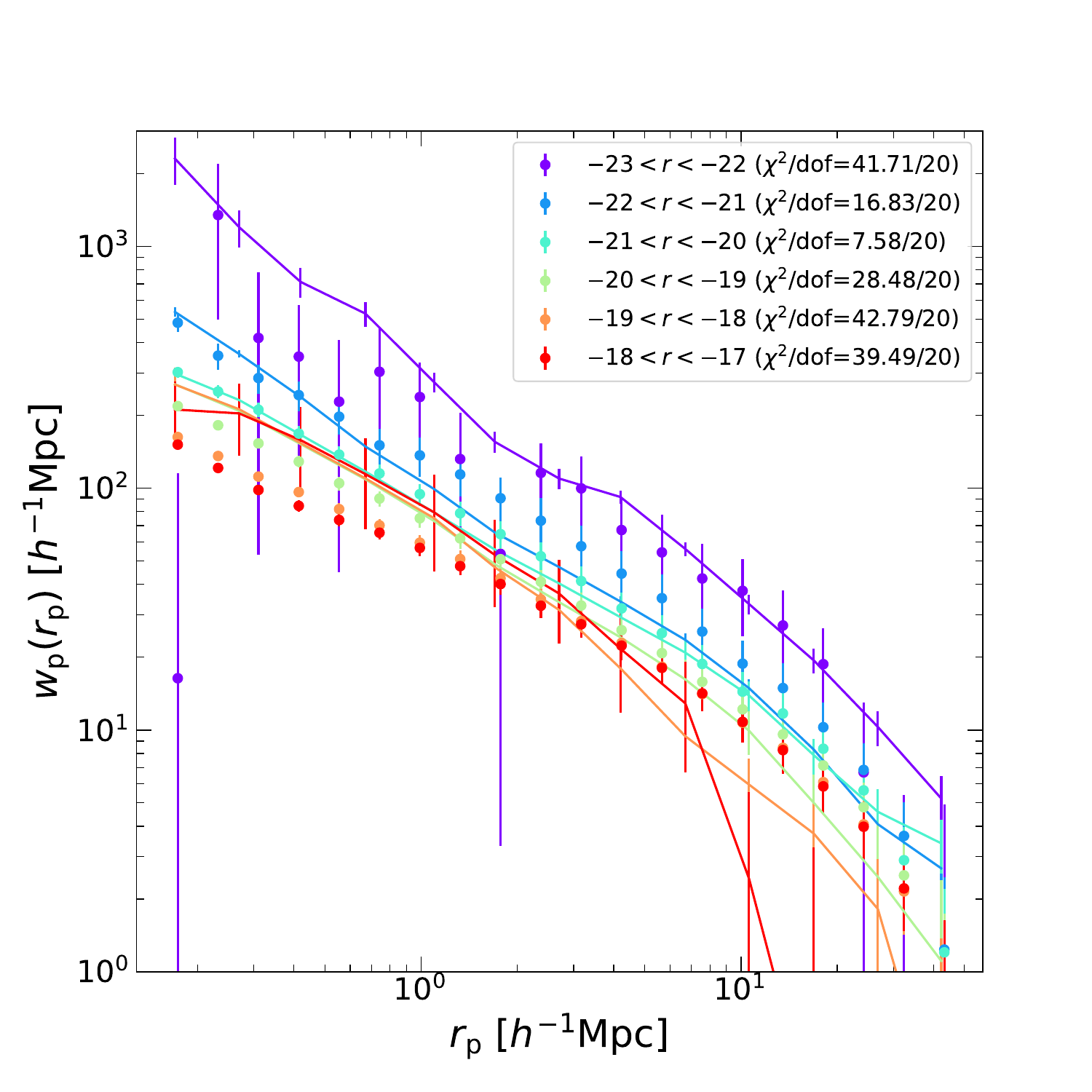}
    \caption{Comparison of the projected two-point correlation function for different luminosity thresholds. The measurements on real observations are represented by solid lines, while the corresponding measurements on the calibrated $\Lambda$CDM mocks are represented by dots. The agreement between observations and mocks is good, in particular on scales involving the 1-halo term (smaller than 1\,$h^{-1}$\,Mpc).}\label{fig:corr_GR}
\end{center}
\end{figure}

\begin{figure}[h!]
\begin{center}
    \includegraphics[scale=0.38]{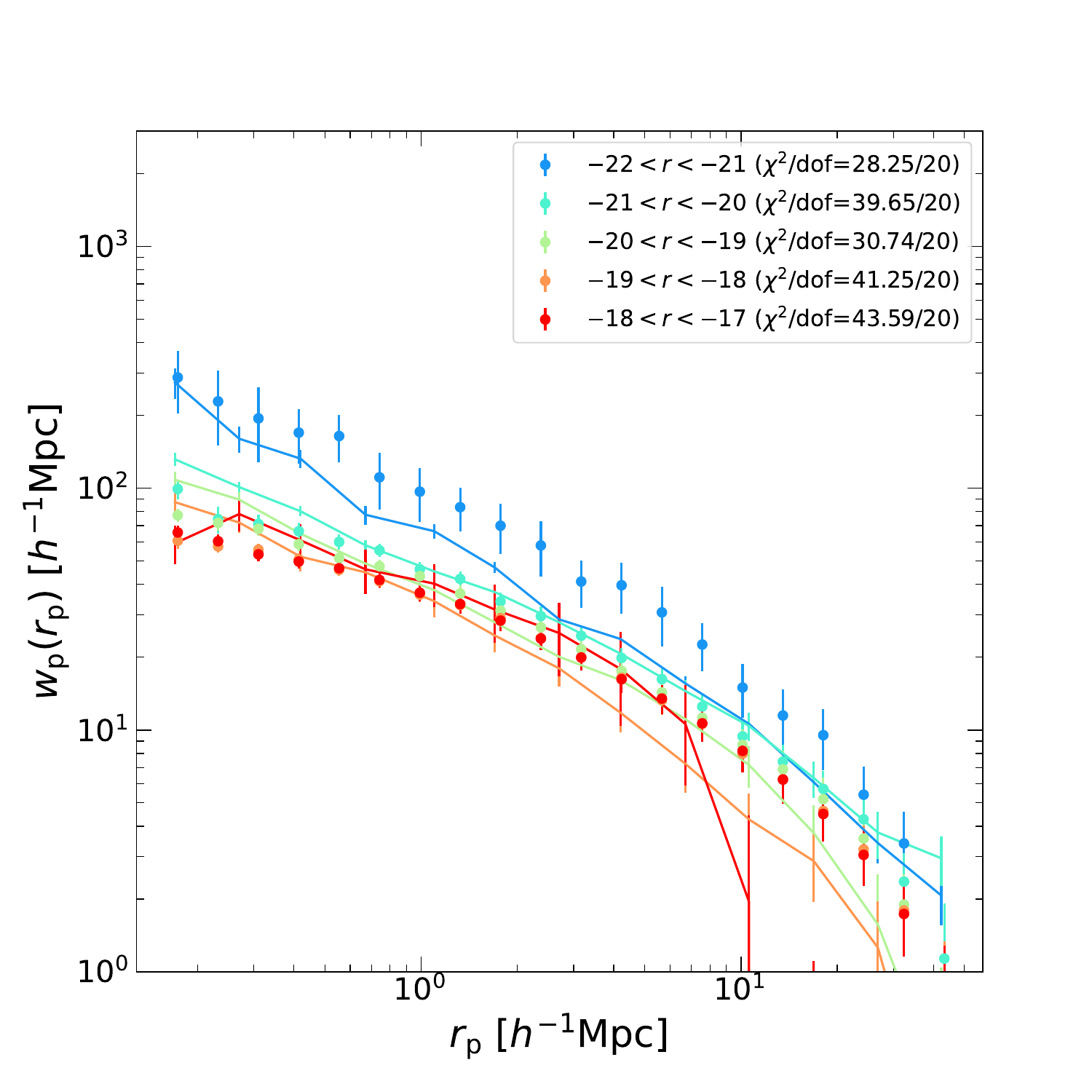}\\
    \includegraphics[scale=0.38]{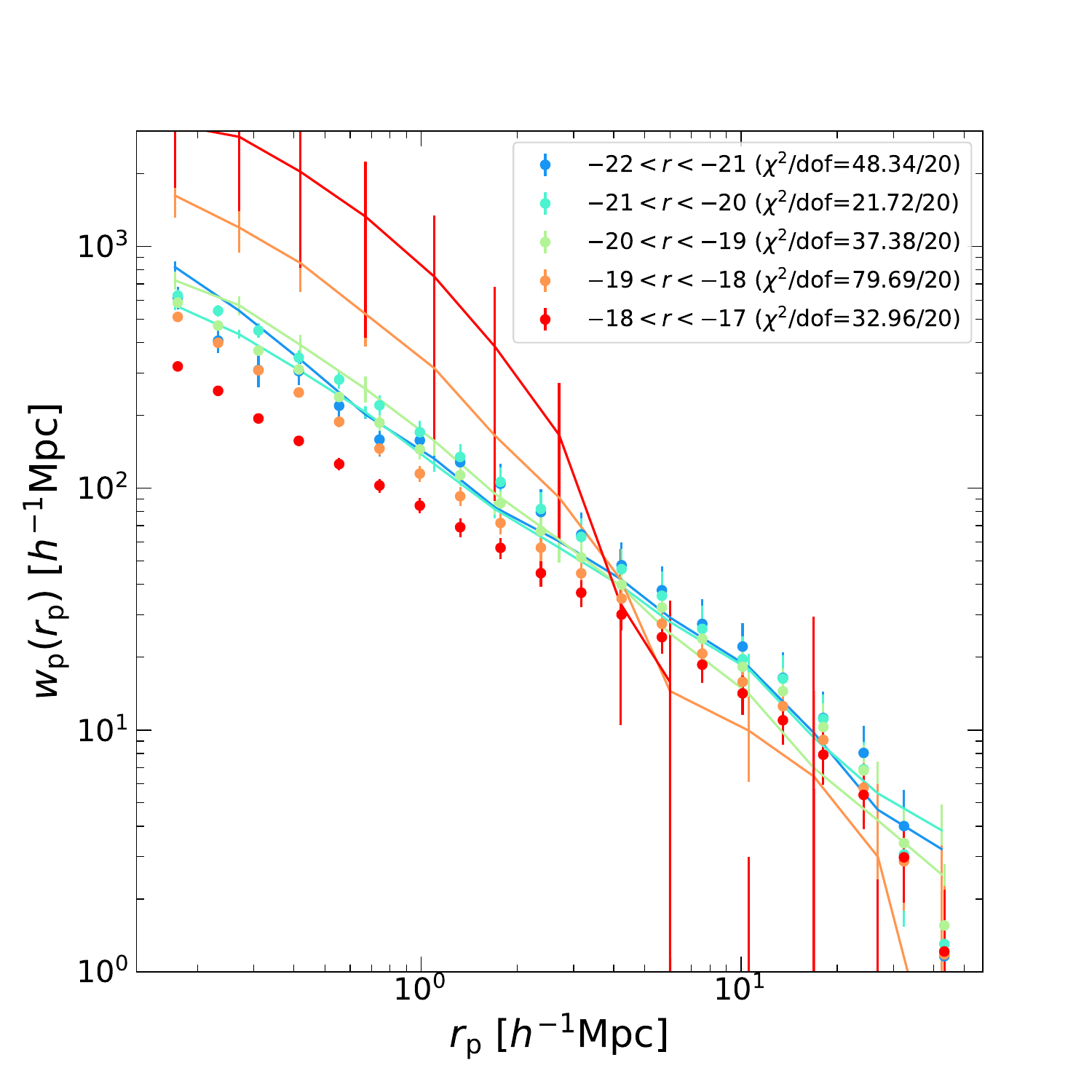}
    \caption{Same as in Fig.\,\ref{fig:corr_GR} but distinguishing between blue (top panel) and red (bottom panel) galaxies.}\label{fig:corr_GR_colors}
\end{center}
\end{figure}

Focussing now on the minimisation part, we fine-tune the hyper parameters of the stochastic optimisation such that the minimisation is reached in a reasonable amount of time. We consider a mutation $M$ randomly selected (at each generation of the population) through a uniform distribution between 0.5 and 1. We set the recombination constant to $R=0.7$, and we use a population of $23$ elements. Our criterion of convergence is fully determined by an absolute tolerance $A=0.0$ and a tolerance $T=0.1$.

In order to start the calibration process, we need to determine the size of the Latyn hyper-cube where the population will be distributed. We note that, since a Latyn hyper-cube in this scenario is just the initial choice to start the sampling, there is no need to optimise it\,\citep[like it happens in the case of building emulators -- see e.g.,][]{EE2}. We only need to ensure that the size of the cube is large enough to contain the minimum. 
We have checked for both halo catalogues that none of the best-fit parameters are close to any edge of the $n$-dimensional cube. Furthermore, we have checked that extending the bounds considered for the different parameters does not change the best-fit obtained.

Although the pipeline is able to automatically stop the calibration when convergence is achieved, we can look at how the $\chi^2$ between the observations and the quantities measured in the mocks evolves. This can be seen in Fig.\,\ref{fig:chi2} for both catalogues. We represent the value of the $\chi^2$ as a function of pipeline evaluations in the form of a two-dimensional histogram. The distributions are very noisy, leading to high values of the $\chi^2$ even after several pipeline evaluations. However, as the number of evaluation increases, we can see how the majority of $\chi^2$ evaluations converge to smaller values. The smallest $\chi^2$ is represented by a red triangle. For $\Lambda$CDM the pipeline has converged after 611 function evaluations, while for the modified gravity mock it has been obtained after 600 function evaluations. We note that given the stochastic nature of the problem, the number of function evaluations might change from one calibration run to another. These values correspond to the final runs presented in this work.

\begin{figure}
\begin{center}
    \includegraphics[scale=0.38]{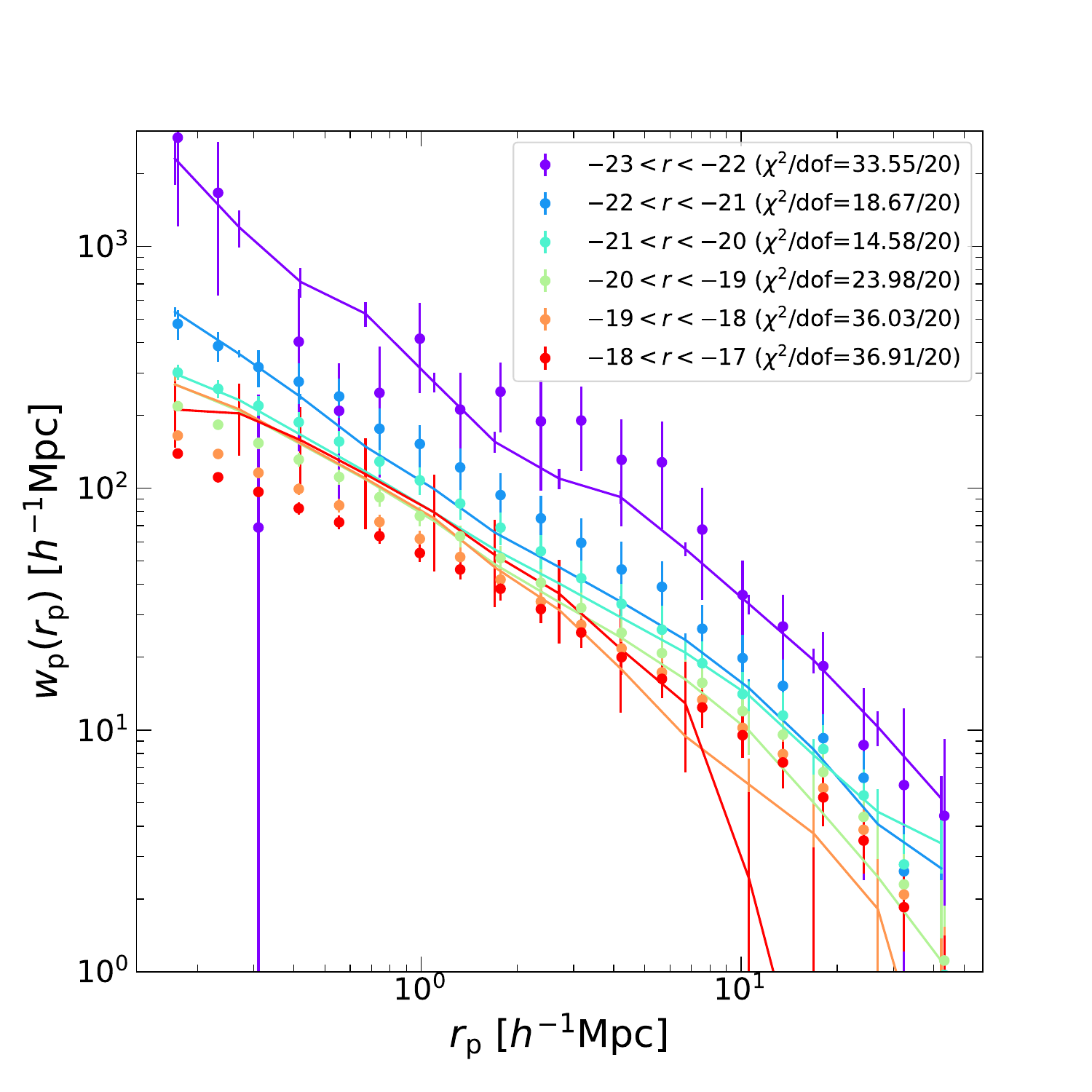}
    \caption{Same as in Fig.\,\ref{fig:corr_GR} but for the $f(R)$ calibrated galaxy mocks.}\label{fig:corr_MG}
\end{center}
\end{figure}

Once the calibration has converged and the mocks have been generated, we can check the final agreement between the measurements from the mocks and the real observations to verify that the calibration worked as expected. Starting with the $\Lambda$CDM case, we show in Fig.\,\ref{fig:corr_GR} the agreement between SDSS projected two-point correlation function measurements for different luminosity thresholds (solid lines) and the corresponding quantities measured in the mock (dots). The agreement is good with $\chi^2$ values that oscillate between 7 and 43 for 20 data points. It is important to note that the agreement is particularly good on small scales (below 1\,$h^{-1}$Mpc), which corresponds to the 1-halo term. The methodology used to populate haloes with galaxies gives us significant freedom to modify the 1-halo term and make it consistent with observations. The situation is slightly more complicated for larger scales, where the clustering is essentially given by the 2-halo term and determined by the clustering of central galaxies and thus that of their host haloes in the input N-body simulation. Therefore, the small excess of clustering for some of the luminosity bins cannot be compensated by a different calibration. In other words, the correlations on large scales are not very sensitive to the galaxy assignment scheme.

In Fig.\,\ref{fig:corr_GR_colors} we show the same comparison but distinguishing between blue (top panel) and red (bottom panel) galaxies. Although in this case the $\chi^2$ values are slightly worse, the agreement is still fairly good; in particular within the 1-halo term. It is also important to mention that we are not considering correlations between different bins (neither in the observations nor in the mocks). Moreover, even after fixing the recipe parameters with the calibration pipeline, the creation of the mock is a stochastic process; therefore, each time a mock is created (with the same parameters) its clustering will be slightly different and the $\chi^2$ values will oscillate. In summary, these $\chi^2$ values should only be interpreted as an indication that a reasonable agreement has been obtained. We also note that when we split between blue and red galaxies we have very few objects in the $-23 < r < -22$ luminosity bin, which prevents us from measuring reliable projected correlation functions. We do not show this luminosity bin in Fig.\,\ref{fig:corr_GR_colors}. 

Looking at red galaxies (bottom plot in Fig.\,\ref{fig:corr_GR_colors}), it is worth mentioning that the clustering amplitude for the low-luminosity bins is lower than the measurements, although the uncertainties on the latter are large. Given the recipe described in Sect.\,\ref{sec:mocks}, there are no ingredients in the model to change the clustering in a colour- and luminosity-dependent way. Moreover, we neglect effects like assembly bias, conformity, or colour segregation insides haloes, for example. However, the goal of this work is to present a pipeline to calibrate the galaxy mock catalogues in an automated way using a realistic recipe for the assignment of galaxy properties, which includes 23 free parameters. Extending the recipe to produce even more realistic galaxy mock catalogues including the above-mentioned effects is beyond the scope of this work.

\begin{figure}
\begin{center}
    \includegraphics[scale=0.38]{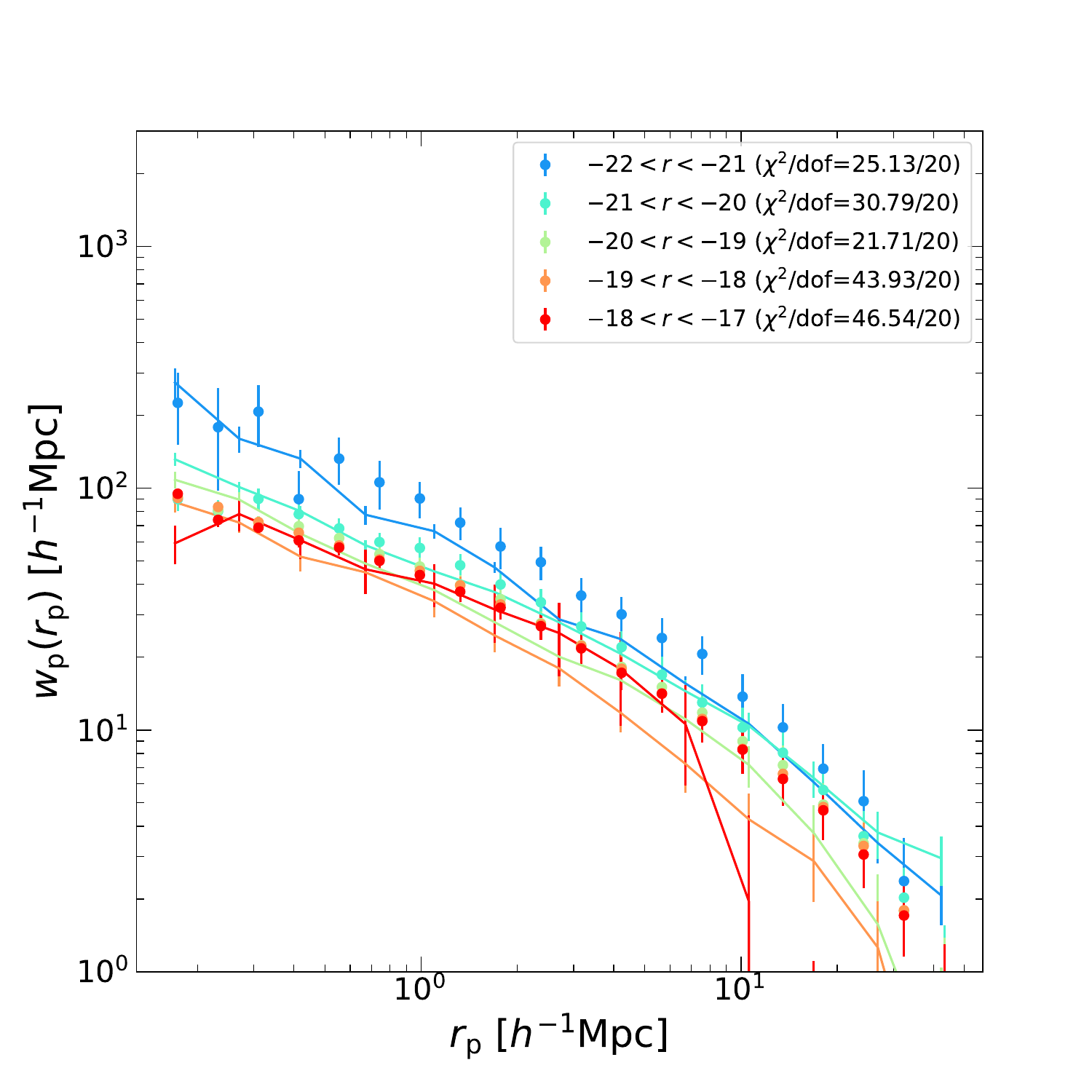}\\
    \includegraphics[scale=0.38]{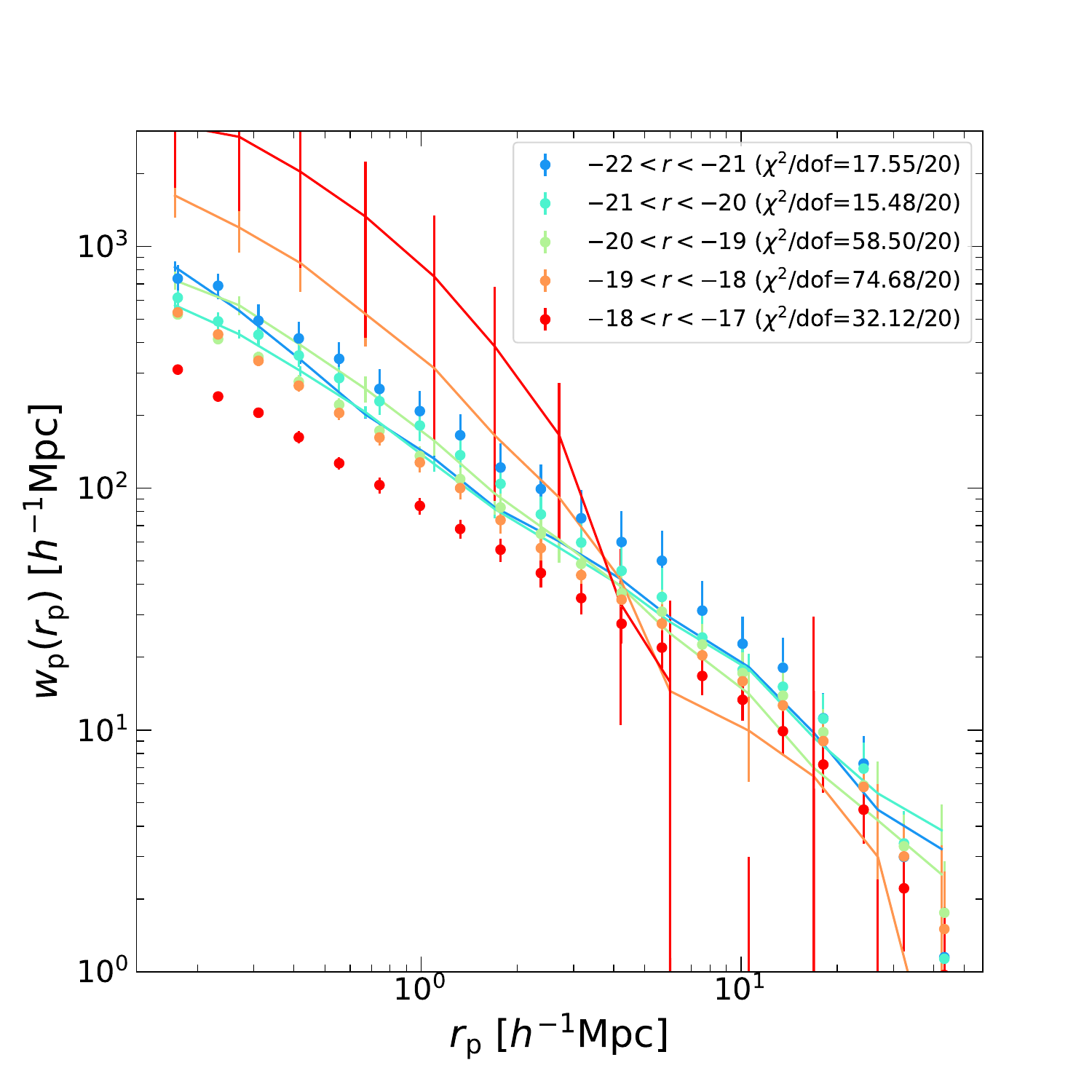}
    \caption{Same as in Fig.\,\ref{fig:corr_GR} but distinguishing between blue (top panel) and red (bottom panel) galaxies and for the $f(R)$ calibrated galaxy mocks.}\label{fig:corr_MG_colors}
\end{center}
\end{figure}

Focussing now on the modified gravity catalogue, we present the comparison to observations in Fig.\,\ref{fig:corr_MG} for all galaxies and Fig.\,\ref{fig:corr_MG_colors} for blue and red galaxies separately. Like in the $\Lambda$CDM case, we do not consider the $-23 < r < -22$ luminosity bin when we do the colour split due to the small amount of objects in the catalogues. The agreement is good, providing $\chi^2$ values between 14 and 37 for 20 data points when we consider all galaxies together. Similarly, the situation is somewhat degraded when splitting between blue and red galaxies. Concerning the latter, we can also observe the lower clustering amplitude in the mock for the low-luminosity bins. However, the overall agreement within the 1-halo term is still fairly good, as expected.

\begin{figure}
\begin{center}
    \includegraphics[scale=0.48]{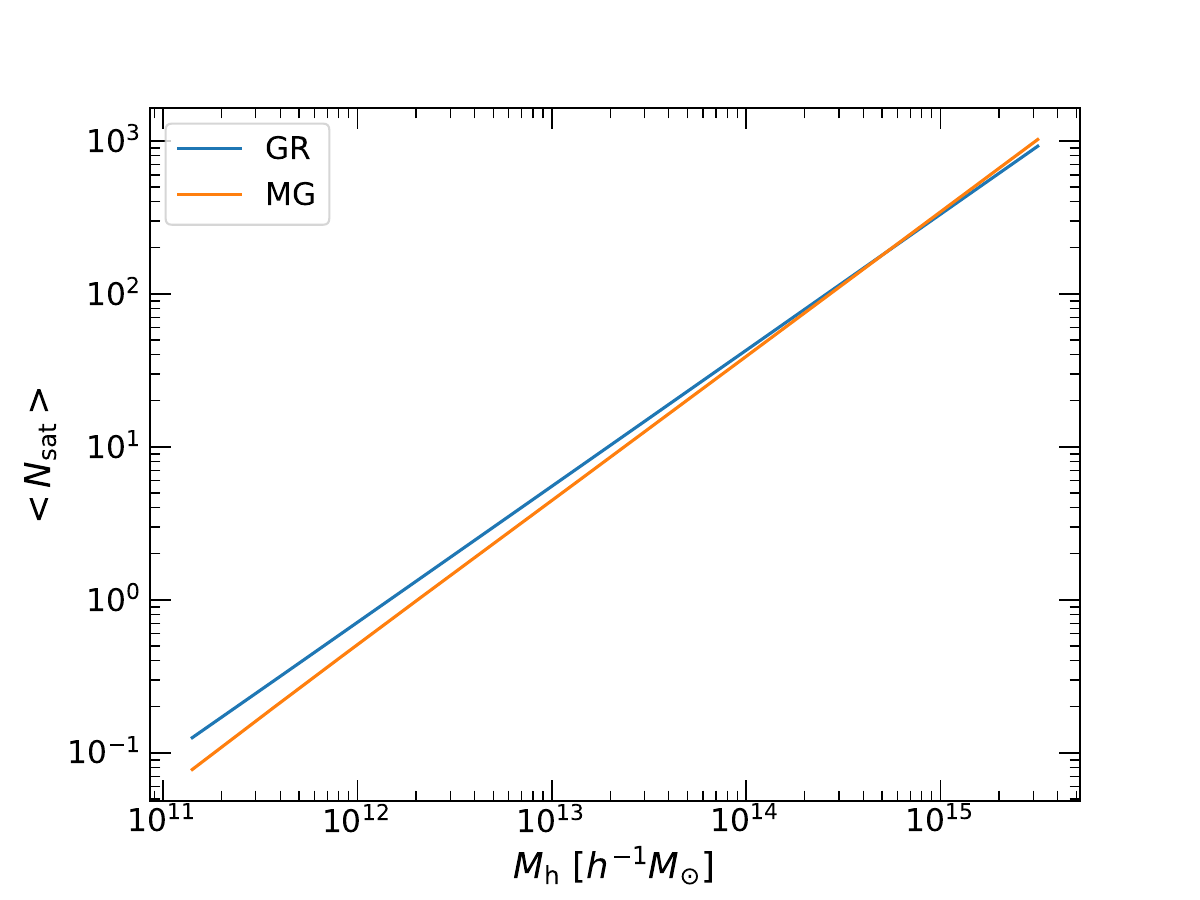}
    \caption{Comparison between the calibrated parameters for the general relativity (GR) and modified gravity (MG) mocks. In this panel we present the impact of different values for $\alpha$, $a_i$, $b_j$, and $s_k$ (see Table\,\ref{tab:table1}) determining the mean of the Poisson distribution to assign the number of satellite galaxies to each halo (see Eq.\,\ref{eq:nsat}).}\label{fig:nsat}
\end{center}
\end{figure}

Although this work is focussed on presenting an automated pipeline to calibrate galaxy mocks against observations, we can also compare the best-fit values obtained for the different parameters of the two cosmologies, for completeness. We can start with $\alpha$, $a_i$, $b_j$, and $s_k$, which correspond to the slope of the satellite mean occupation and how the minimum mass $M_{\rm min}$ depends on the mass of the halo $M_{\rm h}$ (see Table\,\ref{tab:table1} for a summary of the 23 free parameters that have been calibrated). Using these parameters, we can compute the mean of the Poisson distribution used to assign the number of satellite galaxies as a function of the mass of the halo (see Eq.\,\ref{eq:nsat}). We present the evolution of this mean in Fig.\,\ref{fig:nsat} for both cosmologies. We can observe that the modified gravity mock needs fewer satellite galaxies compared to the general relativity case to properly reproduce the observations, except for very massive haloes, where the situation reverses. Concerning the scatter introduced in the abundance-matching relation and the luminosity threshold after which this scatter is applied, the general relativity and modified gravity mocks prefer values that are close between the two cosmologies. 

\begin{figure}
\begin{center}
    \includegraphics[scale=0.48]{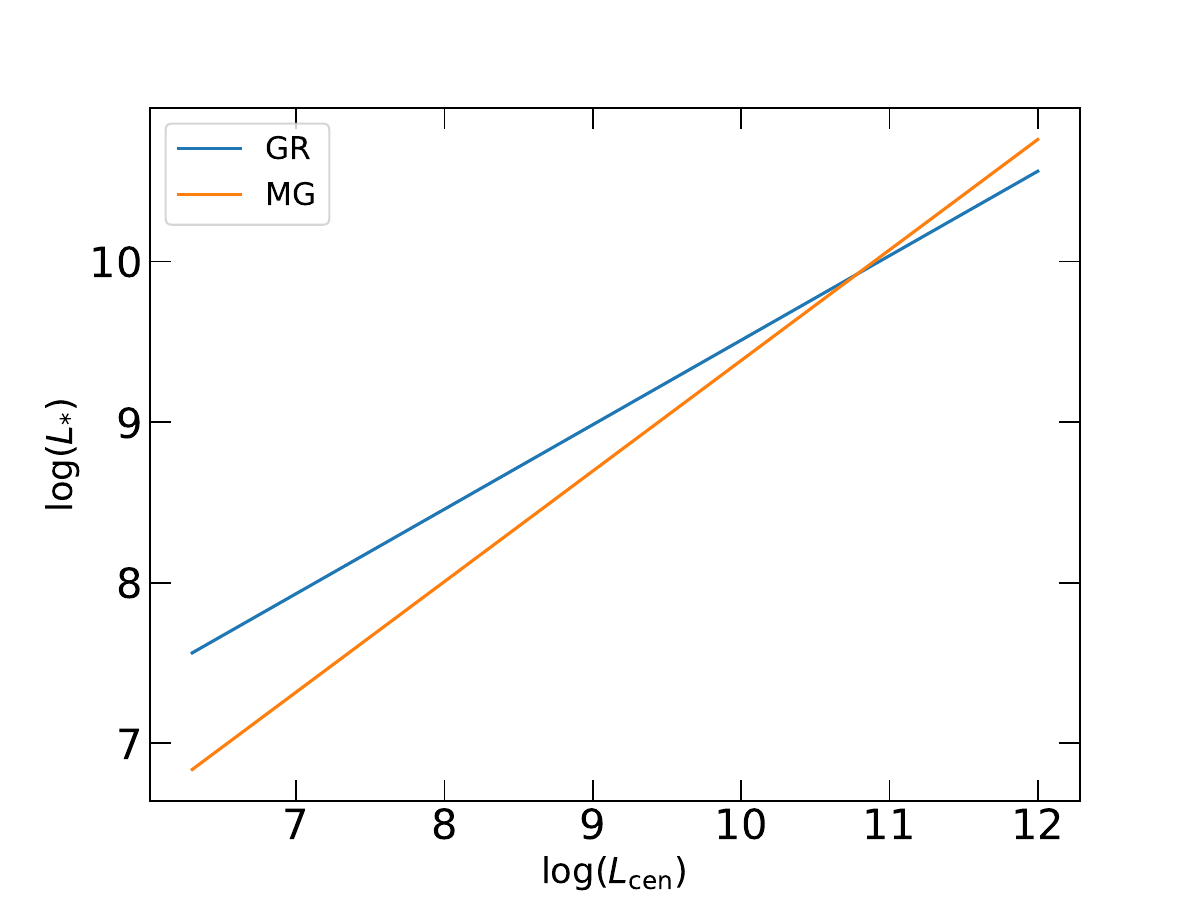}
    \caption{Comparison between the calibrated parameters for the general relativity (GR) and modified gravity (MG) mocks. In this panel we present the impact of different values for $a_{\rm AM}$, $b_{\rm AM}$, and $c_{\rm AM}$ (see Table\,\ref{tab:table1}) determining the relation between the characteristic luminosity and the luminosity of the central galaxy (see Eq.\,\ref{eq:lstar}).}\label{fig:lstar}
\end{center}
\end{figure}

In Fig.\,\ref{fig:lstar} we compare the calibrated values of $a_{\rm AM}$, $b_{\rm AM}$, and $c_{\rm AM}$ through their impact on the relation between the characteristic luminosity and the luminosity of the central galaxy (see Eq.\,\ref{eq:lstar}). We can observe in the figure how the modified gravity mock requires a lower $L_*$ at low $L_{\rm cen}$ values compared to the general relativity case, while the situation reverses at very high $L_{\rm cen}$. Therefore, when we consider low-mass haloes, with fainter central galaxies, the luminosity of satellite galaxies is even lower compared to the central galaxy luminosity in the modified gravity case. This can potentially compensate for the different halo abundance between the modified gravity and general relativity scenarios.

\begin{figure}
\begin{center}
    \includegraphics[scale=0.48]{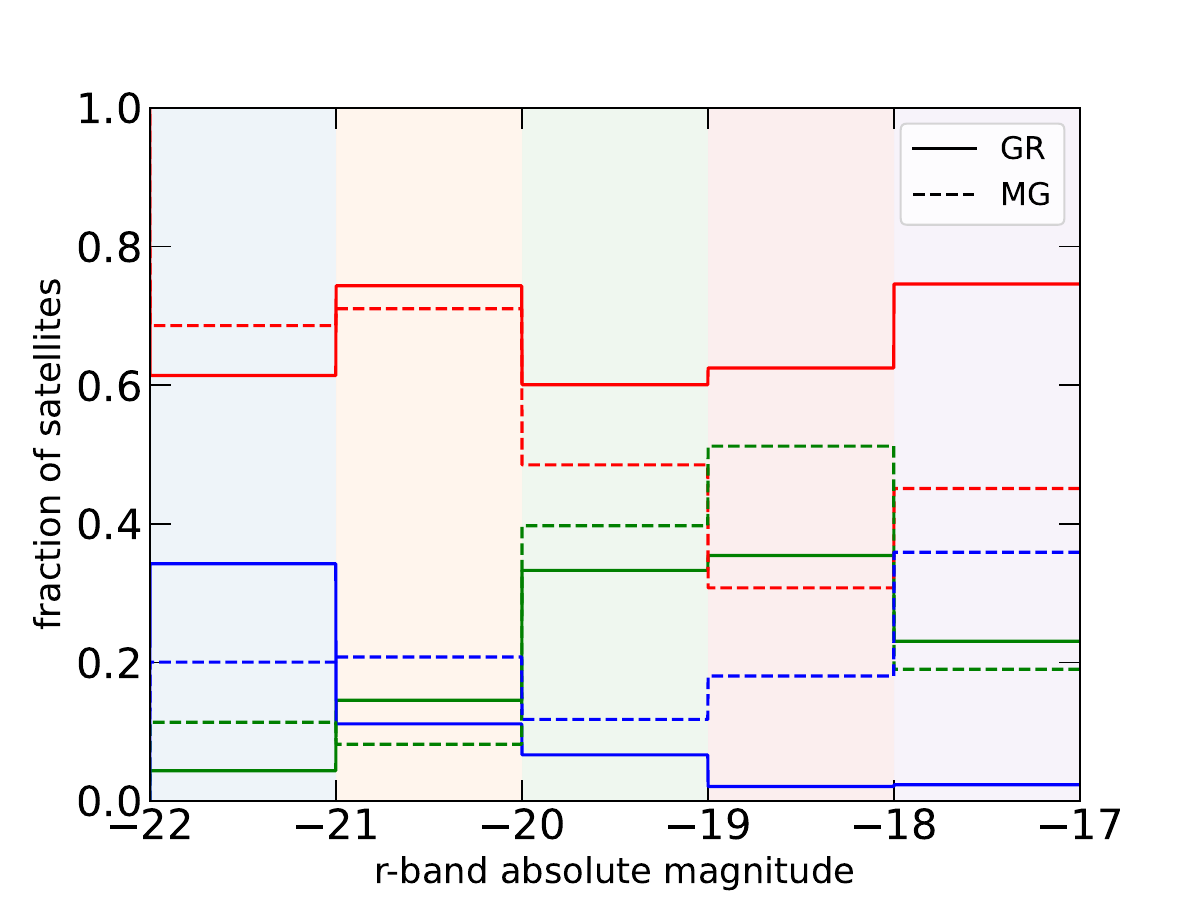}
    \caption{Comparison between the calibrated parameters for the general relativity (GR) and modified gravity (MG) mocks. In this panel we present the evolution of the fraction of satellite galaxies for the different colours and cosmologies.}\label{fig:fsat}
\end{center}
\end{figure}

Finally, we present the fraction of satellite galaxies of each colour for both cosmologies in Fig.\,\ref{fig:fsat}. We can observe that the fraction of red satellites for the general relativity mock remains approximately constant as a function of absolute magnitude, while the modified gravity mock requires fewer red satellites for larger absolute magnitudes. We note that luminosity and colour are correlated. Given the fact that satellite galaxies are fainter in the modified gravity case, it is not unexpected to obtain a smaller number of red satellites, which is needed to keep the colour-magnitude diagram unchanged. Contrary to red satellites, the fraction of green satellites evolves following roughly the same trend for both cosmologies. Given the constraint that the sum of the three colors needs to add up to the total amount of satellite galaxies, we can see the reverse trend for blue satellites, where the modified gravity mock required more blue satellites at larger magnitudes compared to the general relativity case. 

In the figures described above, we have presented a comparison between the different values of the calibrated parameters to understand their changes as a function of cosmology. However, it is important to mention that one would need to have some knowledge of the uncertainty around the best fit before claiming a significant change between the calibrated values of the different parameters. Our goal was only to determine the best-fit values needed to generate a calibrated mock. Therefore, we have only presented a qualitative comparison between the best-fit values for the two cosmologies and leave a more detailed analysis on the physical implications of the exact numerical values for future work.

\section{Conclusions}\label{sec:conclusions}
The use of galaxy mocks has become one of the key ingredients in cosmological analyses. However, current galaxy surveys probe large volumes and small scales, and the next generation of galaxy surveys, like the {\it Euclid} satellite\,\footnote{\url{https://www.euclid-ec.org}}, the Legacy Survey of Space and Time of the Vera C. Rubin Observatory\,\footnote{\url{https://www.lsst.org}}, or the Dark Energy Spectroscopic Instrument (DESI)\,\footnote{\url{https://www.desi.lbl.gov}}, will probe increasingly large volumes with exquisite detail.  
In order to build galaxy mocks that can be used for a large variety of scientific goals, the standard approach is to build massive dark matter N-body simulations and populate the dark matter haloes with galaxies. The connection between dark matter haloes and galaxies have free parameters that need to be tuned to reproduce the real observations. Such fine-tuning becomes extremely difficult when the number of observables, and thus the parameters describing them, increases.  
In this work we provide an automated pipeline to perform such calibration.

We have first described in Sect.\,\ref{sec:mocks} how we model the connection between dark matter haloes and galaxies. We follow closely the combination of HOD and SHAM techniques presented in \citet{Carretero2015}. We have then built a calibration pipeline based on minimising the $\chi^2$ between the quantities observed and the ones measured in the mock. In more detail, we have minimised the discrepancy between the luminosity functions and the galaxy clustering as a function of galaxy colour and galaxy luminosity. Since the process of building a galaxy mock has an inherent stochastic behaviour, we have minimised such $\chi^2$ using the differential evolution algorithm first presented in \citet{Storn1997}. We have generated a population of candidate solutions for the minimum by setting a Latyn hyper-cube in the hyper-parameter space of nuisance parameters to calibrate. We have then randomly selected two members of the population and used their distance to mutate the best candidate solution so far. A new trial population has been built by randomly selecting candidates from the old population or mutated candidates. Using an iterative approach, we have been able to refine our population and stochastically sample the hyper-parameter space to look for its minimum.

Once the calibration pipeline has been prepared, we have applied it to the $\Lambda$CDM and $f(R)$ modified gravity halo catalogues from \citet{Arnold2019}. In both cases the pipeline has been able to converge and provide the minimum of the $\chi^2$ (see Fig.\,\ref{fig:chi2}), therefore setting the values for all the nuisance parameters. We have shown that the best fit found is not close to any of the edges of the hyper-cube and that extending the hyper-dimensional cube does not affect the best-fit obtained. We have also shown that the projected two-point correlation function of the calibrated galaxy mocks is in good agreement with SDSS observations for both the $\Lambda$CDM (see Fig.\,\ref{fig:corr_GR}) and the $f(R)$ (see Fig.\,\ref{fig:corr_MG}) mocks and all the luminosity bins considered. The agreement is particularly good at small scales (below 1\,$h^{-1}$Mpc) where we are well within the 1-halo term. We note that this is the region where our recipes to populate haloes with galaxies have more impact. When we move to larger scales, the clustering is mostly given by the 2-halo term, which is essentially set by the clustering of central galaxies and, therefore, their host dark matter haloes. Changes in our calibration parameters do not have a strong impact on the clustering at large scales. We have also compared the projected two-point correlation function splitting between blue and red galaxies for both $\Lambda$CDM (see Fig.\,\ref{fig:corr_GR_colors}) and $f(R)$ (see Fig.\,\ref{fig:corr_MG_colors}) mocks. The overall agreement between the mocks and the observations is slightly worse in this case, but it is still good within the 1-halo term, where the calibration is more relevant.

To summarise, in this work we have presented a new pipeline able to calibrate galaxy mocks in an automated way. We have applied the pipeline to realistic simulations based on populating halo catalogues with a combination of HOD and SHAM techniques, which leads to a high-dimensional parameter space to calibrate. We have further considered $\Lambda$CDM and modified gravity halo catalogues to show that our calibration pipeline can be applied to extended models. This type of pipelines will be extremely useful for future galaxy mocks, given the large number of parameters that will need to be calibrated. 

Finally, focussing on the modified gravity model, we have created a large, high-resolution galaxy mock for modified gravity that can properly reproduce the real observations at low redshift. In particular, this shows that differences in the cosmological model at low redshift can be re-absorbed by the parameters of the galaxy-halo connection, since both the general relativity and modified gravity simulations reproduce the observations with similar $\chi^2$ values. We now plan to use the galaxy mocks calibrated and generated in this work to explore how future surveys, such as Euclid, LSST, or DESI, can potentially constrain different gravity models \citep[][Alemany-Gotor et al. in prep.]{2025arXiv250404961V}.

\begin{acknowledgements}
PF and FJC acknowledge support form the Spanish Ministerio de Ciencia, Innovaci\'on y Universidades, MCIU/ AEI/10.13039/501100011033/, projects PID2019-11317GB, PID2022-141079NB, PID2022-138896NB; the European Research Executive Agency HORIZON-MSCA-2021-SE-01 Research and Innovation programme under the Marie Skłodowska-Curie grant agreement number 101086388 (LACEGAL) and the programme Unidad de Excelencia Mar\'{\i}a de Maeztu, project CEX2020-001058-M. JC acknowledges support from the grant PID2021-123012NA-C44 funded by MCIN/AEI/ 10.13039/501100011033 and by “ERDF A way of making Europe”. The project that gave rise to these results received the support of a fellowship from "la Caixa" Foundation (ID 100010434). The fellowship code is LCF/BQ/PI23/11970028. This work has made use of CosmoHub, developed by PIC (maintained by IFAE and CIEMAT) in collaboration with ICE-CSIC. It received funding from the Spanish government (grant EQC2021-007479-P funded by MCIN/AEI/10.13039/501100011033), the EU NextGeneration/PRTR (PRTR-C17.I1), and the Generalitat de Catalunya.

\end{acknowledgements}

\bibliographystyle{aa}
\bibliography{example}

\end{document}